Hyperbolic heat conduction, effective temperature and third law in the presence of heat flux

S. L. Sobolev

Institute of Problems of Chemical Physics, Academy of Sciences of Russia, Chernogolovka, Moscow Region, 142432 Russia

E-mail: sobolev@icp.ac.ru

Abstract

Some analogies between different nonequilibrium heat conduction models, particularly, random walk, discrete variable model, and Boltzmann transport equation with the single relaxation time approximation, have been discussed. We show that under an assumption of a finite value of the heat carriers velocity, these models lead to the hyperbolic heat conduction equation and the modified Fourier law with the relaxation term. Corresponding effective temperature and entropy have been introduced and analyzed. It has been demonstrated that the effective temperature, defined as a geometric mean of the kinetic temperatures of the heat carriers moving in opposite directions, is governed by a non-linear relation and acts as a criterion for thermalization. It is shown that when the heat flux tends to its maximum possible value, the effective temperature, heat capacity and local entropy go to zero even at a nonzero equilibrium temperature. This provides a possible generalization of the third law to nonequilibrium situations. Analogies between the effective temperature and some other definitions of temperature in nonequilibrium state, particularly, for active systems, disordered semiconductors under electric field, and adiabatic gas flow, have been shown and discussed. Illustrative examples of the behavior of the effective temperature and entropy during nonequilibrium heat conduction in a monatomic gas, a nano film, and a strong shockwave have been analyzed.

I. INTRODUCTION

Understanding how heat is carried, distributed, stored, and converted in various systems has occupied the minds of many scholars for quite a long time [1-14]. This is not due only to purely academic reasons: its practical importance in the fabrication and characterization of nanoscale systems has been recognized as one of the most critical programs in process industries [15-26]. The presence of the heat flux implies that the system is far from equilibrium. Building a general framework describing the far from equilibrium systems has led to a considerable amount of work towards this aim (Refs.[1-47] and references therein). In spite of the recent advances, our current understanding of the fundamentals of the non-equilibrium heat conduction still remains incomplete, undoubtedly far beyond what we know for equilibrium systems. Strictly speaking, a local temperature has a well-established meaning only in global equilibrium when the heat flux is zero. In particular, the question of what precisely is a "local temperature" in a nonequilibrium system, a concept that has a well-established meaning only in global equilibrium, is open to discussion [5,6,9,14-18,21-24,28-33,40-49]. Classical irreversible thermodynamics (CIT) is based on the local equilibrium assumption, which uses a local temperature defined as in global equilibrium even for nonequilibrium situation with nonzero heat flux. The local equilibrium assumption is valid only for a relatively weak deviation from local equilibrium when the characteristic time scale of the process $t$ significantly exceeds the relaxation time to local equilibrium $\tau$, i.e. $t \gg \tau$. CIT leads to the well-known Fourier law (FL) for the heat flux and parabolic heat conduction equation (PHCE) for the local equilibrium temperature. However, there are two main motivations to go beyond the local equilibrium assumption. One of them, of a theoretical nature, refers to the so-called 'paradox' of propagation of thermal signals with infinite speed, which is predicted by the PHCE [1,2,4,5,6]. The second, more closely related to experimental observations, deals with the propagation of second sound, ballistic phonon propagation, and phonon hydrodynamics in solids at low temperatures, where heat transport departs dramatically from the usual parabolic description [5-7,9-11,14-26]. The most simple and well known modification of the Fourier law (MFL) for the one-dimensional (1-D) case is given by [1,2,5-15]

$$q + \tau \frac{\partial q}{\partial t} = -\lambda \frac{\partial T}{\partial x} \qquad (1)$$

where $q$ is the heat flux, $T$ is the temperature, $\lambda$ is the thermal conductivity. The MFL, Eq.(1), together with the energy conservation law gives the hyperbolic heat conduction equation (HHCE) [1,2,5-15]

$$\frac{\partial T}{\partial t} + \tau \frac{\partial^2 T}{\partial t^2} = a \frac{\partial^2 T}{\partial x^2} \qquad (2)$$

where $a = \lambda/C$ is the thermal diffusivity, and $C$ is the specific heat. The HHCE, Eq.(2), overcomes the paradox of propagation of thermal signals with infinite speed predicting a heat propagation with a finite velocity $v = \sqrt{a/\tau}$. Although Eq.(1) and (2) have been used to describe heat transport for quite a long time, they still raise an important question: how the local nonequilibrium temperature $T$ is defined? Can classical thermodynamic temperature, being an equilibrium concept, still be invoked in the nonequilibrium process described by Eqs.(1) and (2)? The question "what is temperature?" has become a subject of intense theoretical and experimental interest in a more broad context of physics, chemistry and life sciences [5,6,16-18,20-24,27-33,37-47,49]. Several effective non-equilibrium temperatures may be defined, all of which reduce to a common value in equilibrium states, but which yield different results in non-equilibrium situations. For example, in molecular dynamic (MD) simulations, which are often used to study heat flow under far from equilibrium conditions, the most important conceptual problem is how to define the temperature at different planes in the simulation cells. Usually the MD simulations define the temperature $T$ on the bases of an average kinetic energy as [3,5,6,16,18,37-39]

$$\frac{3}{2}k_B T_i = \left\langle \frac{mv_i^2}{2} \right\rangle \quad (3)$$

where $m$ is the mass of an atom, and $v_i$ is the velocity of an atom at site $i$. The temperature defined on the basis of the kinetic energy of the particles is sometimes referred to as the kinetic temperature. The continuous approaches [28,42] also use an analogous definition of local nonequilibrium temperature based on the internal energy – the temperature of the local nonequilibrium state is the temperature of the equilibrium state with the same energy density as in the nonequilibrium state. These approaches assume that the energy density is related to temperature by $u = \int_0^T C d\xi$ and the temperature increase is calculated by $\Delta T = \Delta u / C$ provided that $\Delta T$ is moderate so that there is no phase change and the specific heat can be regarded as a constant [42]. In a more general case the relation between phonon energy and lattice temperature is obtained by Debye model $e(T) = T^4 (9\eta k_B / T_D^3) \int_0^{T_D/T} z^3 (e^z - 1)^{-1} dz$, where $T_D$ is Debye temperature, $\eta = (k_B T_D / \hbar \omega)^3 / 6\pi^2$ is the number density of oscillators [16,27,38]. For glassy systems, the definition of the equilibrium temperature has been extended to the non-equilibrium regime, showing up as an effective quantity in a modified version of the fluctuation–dissipation theorem (FDT) [3,6,17,40]. Glasses are out-of-equilibrium systems in which thermal equilibrium is reached by work exchanged through thermal fluctuations and viscous dissipation exchange that happens at widely different timescales simultaneously. The "active" systems, from phase transformations [12,19,25] to bio systems [31,32,40,47], move actively by consuming energy from internal or external energy sources and their behavior is thus intrinsically out of equilibrium. The effective temperature of the active systems is usually defined also on the basis of the FDT. Extended irreversible thermodynamics (EIT) [5,6,9,21] goes beyond the local equilibrium assumption and obtains generalized heat conduction models by introducing additional state variables, such as heat flux, into the expression of entropy. As a result the nonequilibrium temperature is introduced as $\theta = (\partial S / \partial e)^{-1}$, where $S$ is the local nonequilibrium entropy, *e* is the local energy density. The thermomass (TM) model [41] indicates that the thermal energy is equivalent to a small amount of mass, called thermomass, according to Einstein's mass-energy equivalence relation $E = mc^2$ and modifies the definition of entropy and temperature for nonequilibrium situations. The TM model agrees in many aspects with fluid hydrodynamics [4] and EIT [5,6,9,21].

In this paper we consider a 1D heat conduction when the deviation from local equilibrium is caused by the presence of the heat flux. In Sec.II we briefly review and discuss some different theoretical approaches to transport phenomena to deepen the understanding of heat conduction under far from equilibrium conditions. It has been demonstrated that all these models lead to the MFL and HHCE under an assumption of a finite velocity of heat carriers. Corresponding effective temperature and entropy are introduced and comparisons among different theories are carried out in Sec.III. In Sec.IV we use the results of Sec.III to illustrate the behavior of the effective temperature in some nonequilibrium situations. Concluding remarks are given in Sec.V.

## II. MODELING

### A. Random walk (RW) approach

The ordinary random walk (RW) or Brownian motion is completely characterized by the diffusion coefficient $D \propto h^2/\tau$, where $h$ is the mean free path of the heat (mass) carriers and $\tau$ is the relaxation time. In the limit $h \to 0$ and $\tau \to 0$, the value of the diffusion coefficient is kept nonzero, which, in accordance with the parabolic type of the classical diffusion equation, implies an infinite velocity of diffusion particles $v = h/\tau \to \infty$. For local equilibrium processes with $t \gg \tau$ this physically unpleasant property does not play an important role. However, for relatively fast processes with $t \sim \tau$, a finite value of the particle velocity, which is a more reasonable concept from a physical point of view, should be taken into account. In 1D a well-defined finite velocity of the diffusion particles *v* means that the system consists of two groups of particles – one group moves on the left and another on the right. This two group (TG) approach yields the evolution equations for the particles density as follows [1,2,12]

$$\frac{\partial u_1}{\partial t} + v \frac{\partial u_1}{\partial x} = \frac{u_2 - u_1}{\tau} \quad (4)$$

$$\frac{\partial u_2}{\partial t} - v \frac{\partial u_2}{\partial x} = \frac{u_1 - u_2}{\tau} \quad (5)$$

where $u_1(x,t)$ is density of particles going to the right, $u_2(x,t)$ is density of particles going to the left, $v$ is the velocity of particles, $\tau$ is the mean free time. For following considerations it is convenient to rearrange Eqs.(4) and (5) as follows

$$\frac{\partial u_1}{\partial t} + v\frac{\partial u_1}{\partial x} = -\frac{u_1 - u_0}{\tau_0} \tag{6}$$

$$\frac{\partial u_2}{\partial t} - v\frac{\partial u_2}{\partial x} = -\frac{u_2 - u_0}{\tau_0} \tag{7}$$

where $u_0 = u/2$ with $u = u_1 + u_2$ being the total density of the particles, $\tau_0 = \tau/2$. After some algebra Eqs.(4) and (5) give

$$\frac{\partial u}{\partial t} + \frac{\partial J}{\partial x} = 0 \tag{8}$$

$$J + \frac{\tau}{2}\frac{\partial J}{\partial t} = -\frac{\tau}{2}v^2\frac{\partial u}{\partial x} \tag{9}$$

where $J$ is the particle flux given by

$$J = v(u_1 - u_2) \tag{10}$$

Eq.(10) allows us to represent $u_1$ and $u_2$ in terms of $u$ and $J$ as follows

$$u_1 = (u + J/v)/2 \tag{11}$$

$$u_2 = (u - J/v)/2 \tag{12}$$

Introducing Eq.(7) into Eq.(6), which expresses conservation law in 1D, we obtain

$$\frac{\partial u}{\partial t} + \frac{\tau}{2}\frac{\partial^2 u}{\partial t^2} = \frac{\tau}{2}v^2\frac{\partial^2 u}{\partial x^2} \tag{13}$$

Taking into account that $\tau_0 = \tau/2$ and $D = v^2\tau/2 = v^2\tau_0$, with $D$ being the diffusion coefficient, Eqs.(9) and (13) take the form analogous to Eqs.(1) and (2), respectively. Thus, the assumption of a finite value of the diffusing particle leads to the MFL and the HHCE [1,2].

B. Boltzmann transport equation (BTE)

The BTE with the single relaxation time (or BGK) approximation is given by [6,10,18,24,37,38,42]

$$\frac{\partial f}{\partial t} + \vec{v}\cdot\vec{\nabla}f = -\frac{f - f^0}{\tau_0} \tag{14}$$

where $f$ is the phonon distribution function, $\vec{v}$ is the phonon group velocity, and $f^0$ is the equilibrium distribution function. BTE, Eq.(16), can be cast into an equation for the phonon energy density $e$ by integrating it over the frequency spectrum as $e(T) = \sum_p \int f \hbar \omega_p D_p(\omega) d\omega$, where $p$ is the polarization of phonons (acoustic and optical) and $D_p(\omega)$ is the phonon density of states per unit volume. For simplicity, the effects of temperature on the dispersion relations and the phonon density of states are neglected. Then, the BTE in a phonon energy density ($e$) formulation is given by [37,38,42]

$$\frac{\partial e}{\partial t} + v_x \frac{\partial e}{\partial x} = -\frac{e - e^0}{\tau_0} \quad (15)$$

where $e^0$ is the equilibrium phonon energy density, and $v_x$ is the component of velocity along the $x$-axis. Since in 1D the phonons can travel in the positive or negative direction along the $x$-axis, Eq.(15) gives two equations

$$\frac{\partial e_1}{\partial t} + v \frac{\partial e_1}{\partial x} = -\frac{e_1 - e_1^0}{\tau_0} \quad (16)$$

$$\frac{\partial e_2}{\partial t} - v \frac{\partial e_2}{\partial x} = \frac{e_2 - e_2^0}{\tau_0} \quad (17)$$

Taking into account that $e_i^0 = e/2$, it is evident that Eqs.(16) and (17) have analogous form as Eqs.(6) and (7). Moreover, after some algebra, as above, we obtain equations the energy flux $j$ and energy density $e$:

$$j + \tau_0 \frac{\partial j}{\partial t} = -D \frac{\partial e}{\partial x} \quad (18)$$

$$\frac{\partial e}{\partial t} + \tau_0 \frac{\partial^2 e}{\partial t^2} = D \frac{\partial^2 e}{\partial x^2} \quad (19)$$

where the total phonon energy density is defined as the sum $e = e_1 + e_2$, while the energy flux is given as $j = v(e_1 - e_2)$.

Thus, the BTE with the single relaxation time approximation leads to the constitutive equation for the energy flux $j$, Eq.(18), and the evolution equation for the energy density $e$, Eq.(19), analogous to the MFL, Eq.(1), and the HHCE, Eq.(2).

Note that the transfer equation due to the BTE with the single relaxation time approximation, Eq.(19), is partial differential equation of hyperbolic type. It contains both "relaxation" (or "wave") term $\tau \partial^2 / \partial t^2$ and classical "diffusive" term $\partial^2 / \partial x^2$, so the artificial inclusion of "an additional diffusive term" into the BTE model by Pisipati et al. [38] seems to be excessive.

### C. Lattice Boltzmann method (LBM)

Extensive computational effort is required to solve the BTE , since it involves seven independent variables descriptive for space, time, and momentum or velocity domain. This has led to the development of the lattice Boltzmann method (LBM) that, in essence, is a numerical scheme for solving the BTE, maintaining its accuracy while reducing the computational effort necessary to solve it [37,38,42]. One of the most popular scheme of LBM widely applied in classical phonon hydrodynamics is based on the BTE with the single relaxation time approximation, which, as it has been discussed above, results in the MFL and the HHCE for energy density (temperature). The HHCE describes the space time evolution of the kinetic temperature under the local nonequilibrium conditions when the characteristic time of the process $t\sim\tau$, but the characteristic space scale of the process $L>>h$. This corresponds to the work of Majumdar [24] that obtained the HHCE from semi-classical Boltzmann transport theory only in the acoustically thick limit when the characteristic space scale is much larger than the phonon MFP. Since the LBM is a consequence of the BTE with the single relaxation time approximation and has the same accuracy, it is applicable, strictly speaking, to the local nonequilibrium case with $t\sim\tau$, but is not applicable to the space nonlocal situations when $L\sim h$. This implies that application of the LBM to heat conduction in nano films with $L\sim h$ needs additional justification.

## D. Discrete variable model (DVM)

Although the HHCE overcomes the dilemma of infinite thermal propagation speed of the classical parabolic heat-mass transfer equation, it, as we discussed above, cannot be applied to length scales comparable to the mean free path of energy carriers because of the breakdown of continuum approaches under severe nonequilibrium conditions. Therefore, it is desirable to adopt method directly based on the microscopic view of transport to deal with problems involving both small temporal and spatial scales. This method should also take into account another important issue of nano scale heat conduction - the size of the region over which temperature is defined. The classical definition is entirely local, and one can define a temperature for each space point, whereas for the quantum definition, the length scale is defined by the mean-free-path of the phonon [16]. The idea of the minimum space region to which the local temperature $T(x,t)$ can still be assigned corresponds to the conclusion of Majumdar [24] that "since temperature at a point can be defined only under local thermodynamic equilibrium, a meaningful temperature can be defined only at points separated on an average by the phonon mean free path". It is also consistent with the concept of minimum heat-affected region suggested by Chen [22,23], which assumes that during phonon transport from a nanoscale heat source the minimum size of the heat affected region is of the order of the phonon mean free path.

The most simple approach to overcome the difficulties associated with the nonequilibrium thermal transport at micro/nanoscales is the discrete variable model (DVM) [1,12,13,26,34-36,49], which discretizes the transport process in space and time by defining the minimum lattice size $h$ to which the local temperature $T(x,t)$ can still be assigned and the minimum time $\tau$ (of the order of the mean free time of heat carriers) between the successive events of energy exchange. The DVM temperature cannot vary within a discrete layer on a scale $h$, i.e. one cannot define $T(x,t)$ within this layer because the whole layer is at the same temperature. This point is emphasized, since all theories of heat transport in superlattices have assumed that one could define a local temperature $T(x,t)$ within each layer [16,18]. One might argue that the DVM is analogous to the LBM because both models use the discrete variables. However, as we discussed above, the LBM accuracy is of the order of the accuracy of the BTE with the single relaxation time approximation, which is local in space, whereas the DVM is inherently nonlocal and captures well the behavior of heat transport on short space ($L \sim h$) and time ($t \sim \tau$) scales [26].

The DVM gives the 1D energy transfer equation as follows [1,12,13,26,34-36]

$$U(n+1,k) = \frac{1}{2}[U(n,k+1) + U(n,k-1)] \quad (20)$$

where $U(n,k)$ is the internal energy of a discrete layer $k$ at a discrete time moment $n$. Continuum variables $t$ and $x$ are related with the corresponding discrete variables as follows $t = n\tau$ and $x = kh$. Within a layer $k = x/h$, which in the continuum variables ranges from $(x - \frac{1}{2}h)$ to $(x + \frac{1}{2}h)$, the internal energy $U$ and the corresponding temperature $T$ do not change. In the continuum variables $t$ and $x$, Eq.(20) is given by

$$U(t+\tau,x) = \frac{1}{2}[U(t,x+h) + U(t,x-h)] \quad (21)$$

The discrete formalism implies that the energy exchange between the layers occurs on the border between the neighboring layers $k$ and $k+1$ at an average time moment $(n + \frac{1}{2})$, which gives the following equation for the energy flux $j$ [12,13,26,34-36]

$$j(n+\tfrac{1}{2},k,k+1) = \frac{v}{2}[U(n,k) - U(n,k+1)] \quad (22)$$

Making for convenience a coordinate shift for continuum coordinate $x \to x + h/2$, we can present the heat flux $q$ in terms of the continuum variables as follows

$$j(t+\tau/2,x) = \frac{v}{2}[U(t,x-h/2) - U(t,x+h/2)] \quad (23)$$

where $x$ is a coordinate of the border between the neighboring layers, which centers are at coordinates $x - h/2$ and $x + h/2$. Thus, the DVM is inherently nonlocal – it directly includes into the governing equations for the energy density, Eqs.(20) and (21), and for the heat flux, Eqs.(22) and (23), both time $\tau$ and space $h$ scales of energy carriers.

## 1. Continuum limits

Eqs.(24) and (26) can be represented in an operator form as follows [13,26]

$$[\exp(\tau\partial_t) - \cosh(h\partial_x)]e = 0 \tag{24}$$

$$\exp\left(\frac{\tau}{2}\partial_t\right)q = -\frac{v}{2}\sinh\left(\frac{h}{2}\partial_x\right)e \tag{25}$$

where $e$ and $q$ substitutes for $U$ and $j$, respectively, in the continuum representation. Taylor expansions of these equations in the continuum limit $h\to 0$ and $\tau\to 0$ contain an infinite number of terms with two small parameters $h$ and $\tau$. To obtain the corresponding equations with a finite number of terms one should first specify an invariant of the continuum limit, which conserves a desirable property of the continuum model.

(a) *Diffusive continuum limit* $D = h^2/2\tau = const > 0$. In the continuum limit $h\to 0$ and $\tau\to 0$, Eq.(24) gives up to the first order in $\tau$

$$\tau\frac{\partial e}{\partial t} = \tau D\frac{\partial^2 e}{\partial x^2} + o(\tau)$$

This equation corresponds to the classical heat conduction equation of parabolic type. The requirement that the heat diffusivity $h^2/2\tau$ has a finite value in the continuum limit $h\to 0$ and $\tau\to 0$ implies that the velocity of the heat carriers $v = h/\tau \to \infty$. Indeed, representing $v$ as $v = 2a/h$, we obtain that $v\to\infty$ at $h\to 0$ when $a$ is nonzero. This is the so-called 'paradox' of propagation of energy disturbances with infinite speed discussed above.

(b) *Wave continuum limit* $v = h/\tau = const < \infty$. An alternative type of the continuum limit, which guarantees a finite value of the heat-carrier velocity $v$, requires that $v = h/\tau = const < \infty$ at $h\to 0$ and $\tau\to 0$ [12,13,34-36]. In this case Eq.(24) gives up to the first order in $\tau$

$$\frac{\partial e}{\partial t} + \frac{\tau}{2}\frac{\partial^2 e}{\partial t^2} = \frac{\tau}{2}v^2\frac{\partial^2 e}{\partial x^2} + o(\tau) \tag{26}$$

Eq.(26) is of hyperbolic type and is analogous to Eq.(13) obtained from the RW approach and to Eq.(19) obtained from the BTE with the single relaxation time approximation. Corresponding continuum limit of Eq.(25) gives

$$j + \frac{\tau}{2}\frac{\partial q}{\partial t} = -\frac{\tau}{2}v^2\frac{\partial e}{\partial x} + o(\tau) \tag{27}$$

which also corresponds to the result of the RW, Eq.(9), and the BTE with the single relaxation time approximation, Eq.(18).

*(c) Temperature representation.* Assuming the kinetic definition of the temperature with the constant specific heat $T \propto e/C$, Eqs.(27) and (28) reduce exactly to the HHCE, Eq.(2), and the MFL, Eq.(1), respectively. In terms of the TG picture discussed in the previous sections, the DVM provides the following expressions for the heat flux $q$ (see Eq.(23)) and the kinetic temperature $T$ (see Eq.(21)):

$$q = vC(T_1 - T_2)/2 \qquad (28)$$

$$T = (T_1 + T_2)/2 \qquad (29)$$

where $T_1$ and $T_2$ are the kinetic temperatures of the two group of the heat carries moving in the opposite directions. Eqs.(28) and (29) can be presented in a slightly different form as

$$T_1 = T + q/vC \qquad (30)$$

$$T_2 = T - q/vC \qquad (31)$$

Camacho [27] demonstrates that the two group representation arises due to the Debye approximation in a maximum entropy formalism, which allows one to split the nonequilibrium phonon distribution function in two equilibrium Bose-Einstein distributions for phonons moving to the left and phonons moving to the right, respectively. In the classical limit, the corresponding phonon temperatures are consistent with Eqs.(30) and (31). Kroneberg et al. [28] also assume the TG model and arrive at Eqs.(30) and (31), as well as at the HHCE, Eq.(2), using the energy equations for $T_1$ and $T_2$ analogous to Eqs.(4) and (5). Thus, the DVM with the "wave" law of the continuum limit leads to the HHCE, Eq.(2), and the MFL, Eq.(1).

To conclude this section, it should be noted that the RW [1,2], the TG representation of Kroneberg et al. [28], the BTE with the single relaxation time approximation [28,37,38,42], and the DVM at the wave low of continuum limit [12,13,34-36] lead to the HHCE, Eq.(2), and the MFL, Eq.(1), due to the assumption of the finite value of the heat carriers velocity.

### III. RESULTS AND DISCUSSION

#### A. Effective temperature $\theta$

The kinetic temperature $T$, which space-time evolution is governed by the HHCE, Eq.(2), characterizes the local energy density of the nonequilibrium state – it is equal to the equilibrium temperature of the same system with the same internal energy in equilibrium. In terms of the TG approach it implies that if a local volume element of the nonequilibrium system consisting of the two groups of the heat carriers with the temperatures $T_1$ and $T_2$ is suddenly isolated, i.e. bounded by adiabatic and rigid walls, and allowed to relax to equilibrium, after equilibration the temperature of the local element will be $T = (T_1 + T_2)/2$. However, if the two groups of the heat carriers with $T_1$ and $T_2$ equilibrate reversibly, i.e. while producing work, their common final temperature $\theta$ will be [3,33,49]:

$$\theta = (T_1 T_2)^{1/2}$$

Indeed, before equilibration the total entropy of the two groups is equal to $S_{neq} = k_B \ln T_1 + k_B \ln T_2 = k_B \ln T_1 T_2$, whereas after equilibration $S_{eq} = 2k_B \ln \theta$. The entropy change during the equilibration is $\Delta S = S_{neq} - S_{eq} = k_B \ln T_1 T_2 / \theta^2$. When the system equilibrates reversibly, the entropy does not change ($\Delta S = 0$) and the last expression gives that $\theta$, which will be called as an effective temperature, is equal to the geometric mean of the two temperatures $T_1$ and $T_2$ [33,49].

Multiplying Eq.(30) by Eq.(31), we obtain the following expression for $\theta$ [49]

$$\theta^2 = T^2 - (q/Cv)^2 \qquad (32)$$

For convenience of further discussion, we represent Eq.(32) in the inverse form:

$$T^2 = \theta^2 + (q/Cv)^2 \qquad (32a)$$

Figure 1 shows $\theta$ as a function of the nondimensional heat flux $\hat{q} = q/vCT$ (solid line). In equilibrium $\hat{q} \to 0$ and, as expected, Eq.(32) gives $\theta \to T$. As an absolute value of the heat flux $|q|$ increases, the deviation from equilibrium also increases, which decreases the effective temperature $\theta$. When the heat flux tends to its maximum value $q^{max} = vCT$ (or $\hat{q} = 1$), which is reached when all the heat carriers move in the same direction, Eq.(32) predicts that the effective temperature $\theta$ tends to zero solid line in Fig.1). The limit is also accompanied by $T_1 \to 2T$ and $T_2 \to 0$ (see Eqs.(30) and (31)). Thus Eq.(32) implies, that the real positive values of the effective temperature $\theta$ corresponds to a physically reasonable upper bound on the heat flux $|q| \leq q^{max}$.

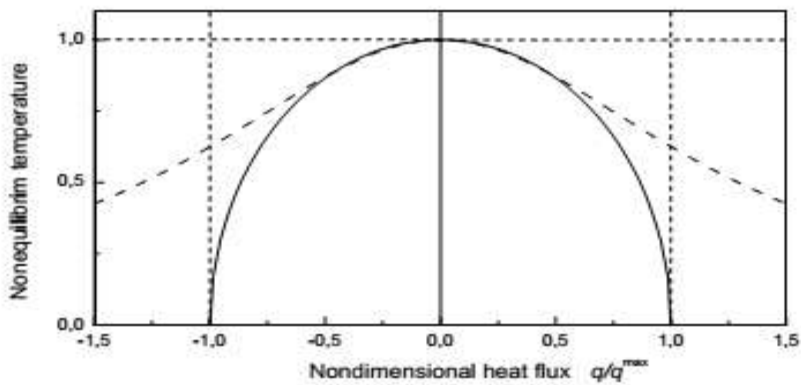

**Fig.1.** Nondimensional effective temperature $\theta/T$ as a function of the nondimensional heat flux $\hat{q}$: solid line – the effective temperature from the present model, Eq.(32); dashed line – the effective temperature from EIT [5,6].

Introducing the stream (drift) velocity $V$ as $V = q/CT$, Eqs.(32) can be rewritten as

$$\theta = T\left(1 - \frac{V^2}{v^2}\right)^{1/2} \tag{33}$$

$$T = \theta\left(1 + \frac{V^2}{v^2}\right)^{1/2} \tag{33a}$$

In equilibrium $V \to 0$ and, as expected, $\theta \to T$. As the heat flux and corresponding stream velocity increase, the effective temperature decreases. Taking into account that $v = h/\tau$ and $D = h^2/2\tau$, the ratio $V/v$ can be presented in terms of the Peclet number $Pe = Vh/D$ as $V/v = Pe/2$.

Note that the factor $\varphi = (1 - V^2/v^2)^{1/2}$ in Eq.(33) arises also as a scaling factor in the effective (thermal) diffusion length $h_{eff} = \varphi h$, which characterizes the effective thermal diffusivity $a_{eff} \propto h_{eff}^2 = a\varphi^2$ ahead of a fast moving heat source [12] or the effective diffusion coefficient $D_{eff} \propto h_{eff}^2 = D\varphi^2$ ahead of a phase transformation zone (for example, during rapid alloy solidification) [19,25]. In these cases the factor $\varphi$ arises due to the form of the HHCE in a moving reference frame [12,19,25]:

$$\frac{\partial T}{\partial t} + \tau \frac{\partial^2 T}{\partial t^2} - V \frac{\partial T}{\partial x} = a\left(1 - \frac{V^2}{v^2}\right)\frac{\partial^2 T}{\partial x^2} + W + \tau \frac{\partial W}{\partial t}$$

where $W$ is the energy source in active systems. The factor $\varphi$ also arises in the different relativistic transformation laws of temperature [5].

The effective heat capacity under the far from equilibrium condition is defined as $C_{neq} = (\partial e/\partial \theta)_q$ [5,6,27]. Using Eq.(32), one obtains

$$C_{neq} = C(1 - \hat{q}^2)^{1/2} \tag{34}$$

*1. Low heat flux limit $|\hat{q}| \ll 1$*

When the deviation from equilibrium is small and $|\hat{q}| \ll 1$, one can expand Eqs.(32) in Taylor series:

$$\theta = T(1 - \hat{q}^2/2) + o(\hat{q}^2) \tag{35}$$

$$T = \theta(1 + \hat{q}^2/2) + o(\hat{q}^2) \tag{35a}$$

The corresponding heat capacity takes the form

$$C_{neq} = C(1 - \hat{q}^2/2) + o(\hat{q}^2)$$

*(a) Interpretation of the effective temperature θ.*

Under the nonequilibrium conditions when $|\hat{q}| > 0$, a part of the kinetic energy used to compute the temperature $T$ is not thermalized. It implies that Eq.(3) for the kinetic temperature $T$ in 1D can be presented as [39]

$$\frac{1}{2}k_B T = \left\langle \frac{m(w_i + V)^2}{2} \right\rangle \tag{36}$$

where $v_i = V + w_i$, with $V$ being the local mean (drift) velocity, and $w_i$ being the thermal randomized velocity of particle $i$, which corresponds to the thermalized kinetic energy. After some algebra Eq.(36) reduces to

$$\frac{1}{2}k_B T = \left\langle \frac{mw_i^2}{2} \right\rangle + \frac{1}{2}mV^2 \tag{37}$$

The fist term on the right hand side of Eq.(37) represents the thermalized (disordered) fraction of the local energy density and can be expressed as $\frac{1}{2}k_B T_{th} = \left\langle \frac{1}{2}mw_i^2 \right\rangle$, where $T_{th}$ corresponds to the temperature of the thermalized (disordered) fraction of the local energy density. Taking into account that $\hat{q} = V/v$, Eq.(37) gives an expression for $T_{th}$ as follows $T_{th} = T(1 - \alpha\hat{q}^2/2)$, where $\alpha = 2v^2 m/Tk_B$. For ideal gas $m/Tk_B = \gamma/c_0^2$ and $v = c_0$ [4], where $c_0$ is sound velocity and $\gamma = C_P/C_V$, which gives $\alpha = 2\gamma$. Thus, comparison of this expression for $T_{th}$ with Eq.(35) allows us to treat the effective temperature $\theta$ as the temperature, which characterizes the thermalized (disordered) fraction of the local energy density (see also discussion in Refs.[5,6]). In other words, the thermal (disordered) fraction of the energy density under local nonequilibrium conditions can be expressed as $e_{th} = \frac{1}{2}k_B\theta = \left\langle \frac{1}{2}mw_i^2 \right\rangle$. The energy of the "ordered" motion of the heat carriers $e_{ord}$, which results in the heat flux $q$, is represented by the difference between the total energy density $e = \frac{1}{2}k_B T$ and the thermal fraction $e_{th}$, i.e. $e_{ord} = \frac{1}{2}k_B(T - \theta)$. Under nonequilibrium conditions $|q| > 0$, the ordered energy $e_{ord} \propto q^2 > 0$, and, consequently, $e(T) > e_{th}(\theta)$ and $T > \theta$. During equilibration the energy of the ordered motion $e_{ord}$ converts into the thermal energy of the disordered motion $e_{th}$, which increases the effective temperature $\theta$. In equilibrium $q = 0$ and, consequently, $e_{ord} = 0$, i..e. the energy of the ordered motion totally transforms into the thermal (disordered) energy, which implies that $e_{th} = e$ and $\theta = T$ (see Fig.1).

*(b) Comparison with gas hydrodynamics.* For small deviation from equilibrium $|\hat{q}| \ll 1$ (or $V/v \ll 1$), Eqs.(33) can be represented as

$$\theta = T\left(1 - \frac{1}{2}\frac{V^2}{v^2}\right) + o(V^2/v^2) \tag{38}$$

$$T = \theta\left(1 + \frac{1}{2}\frac{V^2}{v^2}\right) + o(V^2/v^2) \tag{38a}$$

Bernoulli's equation describing the adiabatic flow of ideal gas is given by [4]

$$T_V = T_0\left(1 - \frac{(\gamma-1)}{2}\frac{V^2}{c_0^2}\right) \tag{39}$$

where $T_V$ is temperature of the flowing gas, $V$ is gas velocity, $T_0$ is gas temperature at $V=0$, $c_0$ is sound velocity at $T_0$. Taking into account that $\theta$, $T$, $V$, and $v$ in the present model correspond to $T_V$, $T_0$, $V$, and $c_0$ in [4], respectively, Eq.(38) and (39) agree fairly well. In fact, the analogy between Eqs.(38) and (39) is a manifestation of the energy conservation law, which allows the energy to transform from the kinetic form of the ordered motion into the thermal energy of the disordered motion.

*(c) Comparison with a maximum entropy formalism.* Following a maximum entropy formalism, Camacho [27] consider a one-dimensional crystal under a heat flux. In the classical limit, Camacho obtains Eq.(35) and concludes that the classical limit condition in nonequilibrium situations becomes a mere generalization of the equilibrium condition where the generalized temperature substitutes the equilibrium temperature.

*(d) Comparison with the EIT.* The EIT [5,6] goes beyond the local equilibrium assumption and obtains generalized heat conduction theory by introducing additional state variables, such as heat flux, into the expression of nonequilibrium entropy. As a result the nonequilibrium temperature $\theta_{EIT}$ is introduced by the EIT as follows [5,6]:

$$\frac{1}{\theta_{EIT}} = \frac{1}{T} + \frac{\hat{q}^2}{2T} \tag{40}$$

To compare this result with the present model, we rearrange Eq.(35) as follows

$$\frac{1}{\theta} = \frac{1}{T} + \frac{\hat{q}^2}{2\theta} \tag{41}$$

Taking into account that for the small deviation from equilibrium $|\hat{q}| \ll 1$ the difference between $\theta$ and $T$ is small and, consequently, the difference between the last terms on the right hand side of Eqs.(40) and (41) is also small, these equation demonstrate fairly good agreement. Fig.1 shows the effective temperature $\theta_{EIT}$ given by the EIT, Eq.(40), as a function of the nondimensional heat flux $\hat{q}$. As it is expected, the effective temperature from the present model $\theta$, Eq.(32), and from the EIT $\theta_{EIT}$, Eq.(41), coincide at a relatively small deviation from equilibrium when $|\hat{q}| \ll 1$, while at a high deviation from equilibrium when $|\hat{q}| \leq 1$, the two temperatures differ substantially (compare solid and dashed curves in Fig.1).

*(c) Comparison with the TM model.* The TM model [41] indicates that the thermal energy is equivalent to a small amount of mass, called thermomass, according to Einstein's mass-energy equivalence relation. In dielectric bulk materials, the thermomass is represented by the phonon gas and the heat transport is thus regarded as the motion of phonon gas with a drift velocity. The momentum balance equation of phonon gas based on gas hydrodynamics [4] gives a generalized heat transport model, which agrees in many aspects with EIT [5,6]. Using the Bernoulli's equation for phonon gas, Dong et al. [41] obtain the relation between the static temperature, $T_{st}$ (effective temperature $\theta$ in the present model), and the total temperature, $T_t$ (kinetic temperature in the present model), which corresponds to Eqs.(38) and (39). For further comparison, we represent the equation for the static temperature $T_{st}$ (Eq.(23) in Ref.[41] ) as follows

$$\frac{1}{T_{st}} = \frac{1}{T_t} + \frac{\hat{q}^2}{T_{st}^2/T_t} \tag{42}$$

The denominators in the last terms on the right hand side of Eqs.(40), (41) and (42) are $\theta$, $T$, and $T_{st}^2/T_t = \theta^2/T$, respectively. It implies that for small deviation from equilibrium when $\theta \approx T$ these equations agree quite well, whereas for high deviation from equilibrium when $\theta$ may be significantly lower than $T$ they differ substantially.

*(d) Nonequilibrium temperature in active systems.* The collective behavior of ''active fluids'', from swimming cells and bacteria colonies, to flocks of birds or fishes, has raised considerable interest over the recent years in the context of nonequilibrium statistical physics [31,32,40]. The active systems consume energy from environment or from internal fuel tanks and dissipate it by carrying out internal movements, which imply that their behavior is more ordered and thus intrinsically out of equilibrium. The energy input in active systems is located on internal units (e.g. motors) and therefore homogeneously distributed in the sample.

Palacci et al. [32] investigated experimentally the nonequilibrium steady state of an active colloidal suspension under gravity field. This work yields a direct measurement of the effective temperature of the active system as a function of the particle activity, on the basis of the fluctuation-dissipation relationship. The effective temperature of the active colloids $T_{eff}$ increases strongly with colloidal activity, which is characterized by the Peclet number $Pe = rV_S/D_0$, where $V_S$ is swimming velocity, $r$ is colloid radius, $D_0$ is equilibrium diffusion coefficient, and is given by [32]

$$T_{eff} = T_b\left(1 + \frac{2}{9}Pe^2\right) \tag{43}$$

where $T_b$ is a bath temperature. The active colloids consume energy from environment in such a way that their motion begins to be more ordered, which increases the effective temperature $T_{eff}$ in comparison with the bath temperature $T_b$. Compared with the present model, the effective temperature $\theta$ is the bath temperature $T_b$, while the kinetic temperature $T$ is the effective temperature of the active colloids, $T_{eff}$. Taking into account that for colloids $\tau = 4r^2/3D$ [32] and $h = 2r$, we obtain that Eq.(38a), expressed in terms of the corresponding Peclet number, gives exactly Eq.(43). Thus, the theoretical prediction of the present model is in a good agreement with the experimental results [32].

Multiple calculations of the effective temperature $T_{eff}$ for self-propelled particles and motorized semi-flexible filaments have been carried out with molecular dynamic simulations by Loi et al. [47] (see also review paper [40]). It has been demonstrated that the FDT allows for the definition of an effective temperature, which is compatible with the results obtained by using a tracer particle as a thermometer [40,47]. It was found that all data can be fitted by the empirical law $T_{eff}/T_b = 1 + \varepsilon f^2$, where $f$ is the active force relative to the mean potential force, $\varepsilon = 15.41$ for filaments and $\varepsilon = 1.18$ for partials [47]. Loi et al. [47] argued that the parameter $f$ plays a role analogous to the Peclet number for colloidal active particles used in the experiments [32]. As well as in the previous case, the effective temperature of active colloids $T_{eff}$ in [47] corresponds to the kinetic temperature $T$ in the present model, while the bath temperature $T_b$ corresponds to the effective temperature $\theta$. This implies that the empirical law obtained by Loi et al. [47] for the effective temperature in active systems is consistent with the present model, Eq.(35a), where the heat flux $\hat{q}$ plays a role of the motor activity $f$.

*2. High heat flux limit $|\hat{q}| \to 1$*

Far from equilibrium, when the high heat flux tends to its maximum value $|\hat{q}| \to 1$, Eqs.(32) and (34) cannot be presented as Taylor's series around $|\hat{q}| = 0$ and the non-linear character of Eq.(32) begins to play an important role. When $|\hat{q}| \to 1$, Eq.(32) and Eq.(34) give that $\theta \to 0$ and $C_{neq} \to 0$, respectively. These results differ substantially from the predictions of the EIT and the TM model (compare solid and dashed lines in Fig.1), which are relevant for low deviation from equilibrium $|\hat{q}| \ll 1$ but agrees with the maximum entropy approach of Camacho [27], who has shown that the high heat flux limit corresponds to the quantum case. Thus, the present model, Eq.(32) and Eq.(34), captures well the behavior of the effective temperature $\theta$ both in the classical limit $|\hat{q}| \ll 1$ and in the quantum limit $|\hat{q}| \to 1$. As we have already mentioned above, the ability of Eq.(32) and Eq.(34) to cover both these limits is a consequence of the analogy between the TG approach and the Bose-Einstein statistics, which is relevant for the quantum limit.

*(a) Disordered semiconductors.* The non-linear relation for the effective temperature has been observed in disordered semiconductors under electric field [5,43-46]. When an electric field is applied to a semiconductor one can characterize the combined effects of the field and the lattice temperature by an effective temperature to describe carrier drift mobility, dark conductivity and photoconductivity [5,43-46]. Marianer and Shklovskii [43] on the basis of their numerical calculations of the liner balance equation for electron transition between localized states in exponential tail have obtained the heuristic formula for the effective temperature

$$T_{eff}^2 = T_0^2 + (Ae_{el}El/k_B)^2 \qquad (44)$$

where $E$ is the electric field, $l$ is the localization length and $e_{el}$ is the electron charge, and $A \approx 0.67$. Baranovskii et al. [44] verified the concept of the effective temperature for the distribution of electrons in band tails under the influence of a high electric field using a new Monte-Carlo simulation algorithm. The simulated data demonstrated a good agreement with the phenomenological equation (44) in a wide temperature range 3<$T$<150 K. These results indicate, that the concept of the effective temperature can in fact be used as a substitute for the combined action of both the applied electric field and the temperature, as far as relaxation processes are concerned [44]. Nebel et al. [45], who experimentally measured the electric-field-dependent dc dark conductivity over a broad temperature range (10<$T$<300 K) in phosphorus- and boron-doped and intrinsic amorphous hydrogenated silicon (a-Si:H), found a good agreement with the phenomenological expression, Eq.(44). Liu and Soonpaa [46] experimentally demonstrated the similarity between temperature and electric-field effects in thin crystals of $Bi_{14}Te_{11}S_{10}$ and observed a good agreement with Eq.(44), particularly at low temperatures from $T$=1.8 to 4.5 K. Liu and Soonpaa [46] noted that the quantum effects played an important role in their experiments due to the samples size of five atoms thick and the low temperatures.

Compared with the present model, the effective temperature $\theta$ is the crystal temperature with zero electric field $T_0$, while the kinetic temperature $T$ is the effective temperature of the crystal under electric field $T_{eff}$. Taking into account that the electric current $i = \sigma_E E$, where $\sigma_E$ is the electrical conductivity, plays a analogous role as the heat flux $q$ (see, for example, Ref.[6]), we obtain that Eq.(44) corresponds to Eq.(32a). Note that although the heuristic Eq.(44) provides a good comparison with the experimental data [44,45] and is helpful from a practical point of view, it did not obtained a physical interpretation [5,45].

More recently, Pachoud et al. [48] experimentally investigated electron transport in granular graphene films self-assembled by hydrogenation of suspended graphene. The authors measured the conductance $G$ of different bias voltages $U$ and temperatures $T$ to extract the typical localization length of the samples $l$ at different temperatures between 2.3 K and 20 K. It was shown that charge carriers experience an effective temperature $T_{eff}$, which is described by Eq.(44). Importantly, $T_{eff}$ uniquely determines $G$, which implies that constant-conductance domains of $(U^2, T^2)$ - space are straight lines of slope $-(Ae_{el}l/L_{ch}k_B)^2$, where $L_{ch}$ is the channel length and $L_{ch} = U/E$ [48]. It has been also demonstrated that two different regimes can be clearly distinguished in the behavior of the standard deviations $\sigma_{\ln G}$ of the log-conductance as a function of $T_{eff}$: below $T_{eff} = 10$ K, $\sigma_{\ln G}$ is weakly temperature-dependent while above 10 K, $\sigma_{\ln G}$ decreases rapidly with $T_{eff}$. This implies that the concept of the effective temperature is very useful for analyzing transport phenomena in the granular graphene materials [48].

Thus, the theoretical prediction for the non-linear definition of the effective temperature, Eq.(32), is in good agreement with the experimental results [45,46,48]. Remarkable that this agreement holds in a wide temperature range up to very low temperatures where the quantum effects begin to play an important role.

### 3. Some comments

*(a) Space time evolution of the effective temperature.* The space-time evolution of the effective temperature $\theta(x,t)$ can be calculated by two ways. First way is to calculate $T(x,t)$ and $q(x,t)$ using the HHCE, Eq.(2), and the MFL, Eq.(1) and, then, to calculate $\theta(x,t)$ using Eq.(32a). Another way is to calculate $T_1(x,t)$ and $T_2(x,t)$ using the HHCE and then calculate $\theta$ from the relation $\theta = (T_1 T_2)^{1/2}$. Note that although $T(x,t)$, $T_1(x,t)$ and $T_2(x,t)$ are governed by the same HHCE, Eq.(2), they do not coincide due to the different corresponding boundary and/or initial conditions.

*(b) Effective and reference temperatures.* Summarizing this section, we would like to comment on possible inversion of the effective and reference temperatures. The active systems are out of equilibrium due to the consumed energy from environment. In this case, the (nonequilibrium) temperature of the active system $T_{eff}$ depends on the motor and plays a role of the effective temperature, while the ambient (equilibrium) bath temperature $T_b$ plays a role of the reference temperature. In relaxing systems, which are initially out of equilibrium and relax to equilibrium without consuming energy, the effective temperature $\theta$ acts as a criterion for thermalization, i.e. characterizes the thermal (equilibrated) fraction of the internal energy and, in this sense, is analogous to the bath (equilibrium) temperature $T_b$. The kinetic temperature in relaxing systems $T$, as well as the effective temperature in active systems $T_{eff}$, characterizes the total energy density of the nonequilibrium state. In equilibrium $T_{eff} = T_b$ and $T = \theta$. However, the effective temperature in active systems $T_{eff}$ increases with increasing motor activity due to the consumed energy, and, consequently, always $T_{eff} \geq T_b$ (see Eq.(35a)), whereas in relaxing systems always $\theta \leq T$ (see Eq.(35)), where the kinetic temperature $T$ plays a role of a reference temperature. Therefore, it is important not to be confused concerning the definitions of the different effective and reference temperatures under far from equilibrium conditions (see also discussion in Ref.[5]).

## B. Effective entropy

The information entropy is given by [3]

$$S = -\sum_i u_i \ln u_i$$

where $u_i$ is the distribution function of subsystem $i$. For the system under consideration we have two subsystems ($i=1,2$), which distribution function can be represented in terms of the corresponding temperatures as $u_i = T_i / 2T$. In such a case this equation takes the form

$$S = -[T_1 \ln(T_1/2T) + T_2 \ln(T_2/2T)] / 2T \tag{45}$$

Using Eqs.(30) and (31) for $T_i$, the expression for entropy, Eq.(45), can be rewritten in terms of heat flux as

$$S = \ln 2 - \frac{1}{2}(1+\hat{q})\ln(1+\hat{q}) - \frac{1}{2}(1-\hat{q})\ln(1-\hat{q}) \tag{46}$$

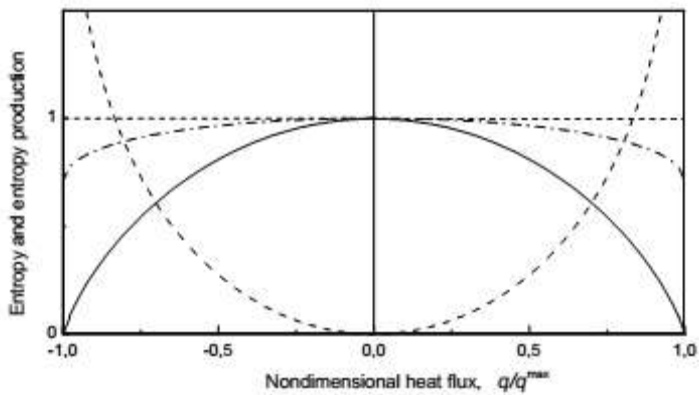

**Fig.2.** Nonequilibrium entropy $S$, Eq.(46), scaled with $S_{eq}$, (solid line) and the entropy production $\sigma_S$, Eq.(47), (dashed line) as functions of the nondimensional heat flux $\hat{q}$. The nonequilibrium entropy obtained by Camacho [27] from a maximum entropy formalism is placed for comparison (dash-dotted line).

The nonequilibrium entropy $S$, Eq.(46), scaled with $S^{eq}$, is shown in Fig.2 as a function of the nondimensional heat flux $\hat{q}$ (solid line). As expected, $S$ is always less than or equal to that of a local equilibrium situation $S^{eq} = \ln 2$. The presence of the heat flux reduces the value of $S$, indicating that the nonequilibrium state is more ordered than for the corresponding equilibrium state.

The Lagrange multiplier $\gamma$ assigned to the heat flux constraint, can be calculated as

$$\gamma = \left(\frac{\partial S}{\partial q}\right)_e = -\frac{1}{2}\ln\frac{1+\hat{q}}{1-\hat{q}}$$

The parameter $\gamma$ has no analog in equilibrium and must be regarded as a purely nonequilibrium quantity describing how an increment in the heat flux modifies the entropy [6,27]. Fig.3 shows minus $\gamma$ as a function of $\hat{q}$.

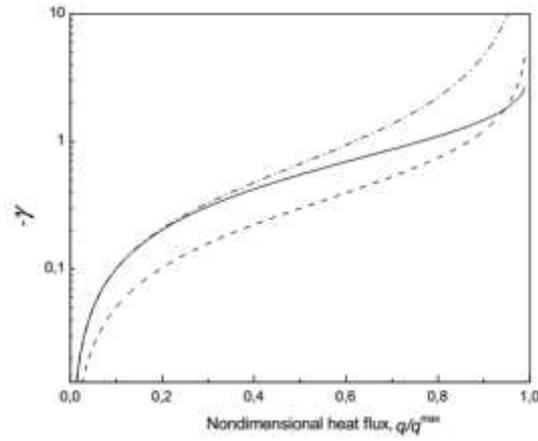

**Fig. 3.** Parameter minus $\gamma$ as a function of the nondimensional heat flux $\hat{q}$: solid line – the present model, dashed line – the quantum limit by Camacho [27], dash-dotted line – the classical limit by Camacho [27].

To introduce the corresponding entropy production $\sigma_S$, let us consider, following the EIT [5,6], a volume element which is sufficiently small so that within it the spatial variation of temperature is negligible. If the volume element is suddenly isolated and allowed to decay to equilibrium, the entropy production would be $\sigma_S = \dot{S} = \gamma \partial \hat{q}/\partial t'$, where $t' = t/\tau$ is the nondimensional time. Taking into account that for the small volume element Eq.(2) gives $\partial \hat{q}/\partial t' = -\hat{q}$, the entropy production can be expressed as

$$\sigma_S = -\frac{\hat{q}}{2}\ln\frac{1+\hat{q}}{1-\hat{q}} \qquad (47)$$

Fig.2 shows $\sigma_S$ as a function of $\hat{q}$ (dashed line). In equilibrium ($\hat{q}=0$) and, as expected, $\sigma_S = 0$. When $\hat{q} \to 1$, Eq.(47) gives $\sigma_S \to \infty$.

*1. Low heat flux limit $|\hat{q}| \ll 1$*

For small deviation from equilibrium ($\hat{q} \ll 1$), the expression for the entropy $S$, Eq.(46), and the entropy production $\sigma_S$, Eq.(47), can be expressed as

$$S = S^{eq} - \hat{q}^2/2 + o(\hat{q}^2)$$
$$\sigma_S = \hat{q}^2 + o(\hat{q}^2)$$

which agree with the expression for the local nonequilibrium entropy and entropy production obtained by Jou et al. [5,6] in the framework of the EIT and by Dong et al [41] in the framework of the TM model. In the limit the parameter $\gamma$ reduces to $\gamma = -\hat{q} + o(\hat{q})$, which corresponds to the classical limit by Camacho [27] (compare solid and dash-dotted lines in Fig.3).

### 2. High heat flux limit $|\hat{q}| \leq 1$

At high deviation from equilibrium when $|q| \to q^{max}$, Eq.(46) for the nonequilibrium entropy $S$ results in $S \to 0$ (see solid line in Fig.2). This can be understood microscopically as follows: as the heat flux grows, the number of heat carriers moving contrary to the heat flow decreases, and in the limit $|q| \to q^{max}$ they disappear. In such a case all the heat carriers move in the same direction with the same velocity, which is completely ordered state with $S = 0$. At the same, the entropy production $\sigma_S$, Eq.(47), tends to infinity (see dashed line in Fig.2).

Thus, when the heat flux $q$ tends to its upper limit $|q| \to q^{max}$, Eqs.(33), (34), and (46) predict that $\theta \to 0, C_{neq} \to 0$, and $S \to 0$. This provides a generalization of the third law to the far from equilibrium situation: indeed, in equilibrium, $\theta$ coincides with the kinetic temperature $T$, however, in nonequilibrium, we have $\theta \to 0, C_{neq} \to 0$, and $S \to 0$ at $|q| \to q^{max}$ even at a non-zero value of $T$ (see also discussion about the third law in Refs.[5,6,27]).

## IV. ILLUSTRATIVE EXAMPLES

### A. Effective temperature in monatomic ideal gas

Let us consider a virtual relaxation to local equilibrium of a small adiabatically isolated system where the heat carriers are placed uniformly and move in the same direction. In other words, the initial condition for the situation is: $q = q^{max}$ at the initial time moment $t=0$ (see Fig.4a). In this case Eqs.(30) and (31) give the initial conditions for the temperatures $T_1$ and $T_2$ as follows: $T_1 = 2T$ and $T_2 = 0$, while Eq.(32) gives $\theta = 0$. As we discussed above, the heat flux in the system is governed by the equation $\partial \hat{q}/\partial t' = -\hat{q}$, which gives $\hat{q}(t) = \exp(-t/\tau)$ (see also [5,6,]).

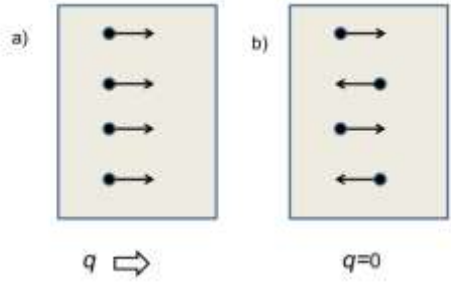

**Fig.4** a) Schematic representation of the nonequilibrium state with the maximum heat flux $q = q^{\max}$ when all the heat carriers move in the same direction. In this case $T_1 = 2T$, $T_2 = 0$, and $\theta = 0$; b) Schematic representation of the equilibrium state with $q = 0$. In this case $T_1 = T_2 = \theta = T$.

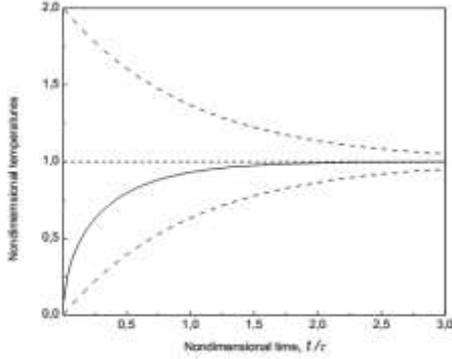

**Fig.5.** Nondimensional effective temperature $\theta/T$ (solid line) as a function of nondimensional time $t/\tau$ during relaxation from the nonequilibrium state (see Fig.4a) to the equilibrium state (see Fig.4b). The temperatures $T_1/T$ (upper dashed line) and $T_2/T$ (bottom dashed line) are also shown for comparison.

Accordingly, the temperatures, $T_1$ and $T_2$, tend to the equilibrium temperature $T$ as $T_1 = T[1 + \exp(-t/\tau)]$ and $T_2 = T[1 - \exp(-t/\tau)]$, respectively (see dashed lines in Fig.5). The effective temperature $\theta$, Eq.(32), increases from zero at $t = 0$ to its maximum value $\theta^{\max} = T$ in the equilibrium state at $t \to \infty$ (see solid line in Fig.5).

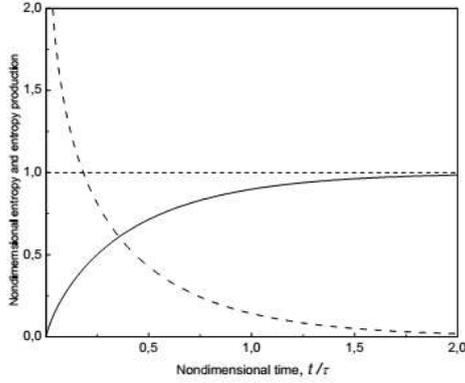

**Fig.6.** Nonequilibrium entropy $S$, Eq.(46), scaled with $S_{eq}$, (solid line) and the entropy production $\sigma_S$, Eq.(47), (dashed line) as functions of as functions of nondimensional time $t/\tau$ during relaxation from the nonequilibrium state (see Fig.4a) to the equilibrium state (see Fig.4b).

Fig.6 shows the time evolution of the nonequilibrium entropy $S$, Eq.(46), scaled with $S^{eq}$, and the corresponding entropy production $\sigma_S$ (solid and dashed lines, respectively). As expected, $S/S^{eq}$ increases from zero at $t=0$ to unity at the equilibrium state at $t \to \infty$ (solid line in Fig.6). Accordingly, $\sigma_S$ decreases from infinity in the initial nonequilibrium state at $t=0$ to zero in the equilibrium state at $t \to \infty$ (dashed line in Fig.6)

The time evolution of $T_1$ and $T_2$ is analogous to the behavior of the effective temperatures for a birth-death process in gene networks [32]: as the coupling strength between species increases, the effective temperatures of the species tend to equalize, as the ''hotter'' one drops and the ''cooler'' one increases reaching the average temperature (compare Fig.5 in the present paper with Fig.4a in Ref.[32]).

Now let us compare the behavior of $T_1$, $T_2$, and $\theta$ in the present model (Fig.5) with the LBM simulation of pico- and femto-second laser heating of silicon [42]. In spite of the fact that the LBM simulation calculates the temperature distribution in the bulk silicon as functions of coordinate, whereas the present model gives the temperatures as function of time, the results can be qualitatively compared because they both consider the energy evolution due to interaction (relaxation) between different modes. So, after the laser heating in LBM simulation [42] stops, the equivalent temperature in the laser incidence direction, which corresponds to $T_2$ in the present paper, decreases with coordinate, while the equivalent temperature in the opposite direction, which corresponds to $T_1$ in the present model – increases. This behavior exactly corresponds to the time evolution of $T_2$ and $T_1$ (see Fig.5). Moreover, the equivalent temperature in the LBM simulation [42], associated with the energy flowing in one of the lateral directions, increases with coordinate in analogy to the increase of the effective temperature $\theta$ in time (see solid curve in Fig.5). Both temperatures increase due to equalization of the initially non-uniform distribution of energy between different degrease of freedom.

### B. Steady-state heat conduction in a nano film

#### 1. Effective transport coefficients

Recently, the trend towards miniaturization of electronic devices has increased the interest in nonequilibrium effects during nano-scale heat conduction. The DVM gives the effective thermal conductivities across a nano film as follows [26]

$$\frac{\lambda_{\pm}^{eff}}{\lambda} = \frac{1}{1 \pm h/L} \qquad (48)$$

where $\lambda$ is the bulk thermal conductivity. The sign "+" corresponds to the thermal conductivity with allowance for the temperature jump at the boundaries between the thermal reservoirs and the film, whereas the sign "−" corresponds to the effective thermal conductivity, which is based on the temperature gradient inside the film and does not take into account the temperature jumps at the boundaries [26]. In the Fourier regime $L \gg h$ or $Kn \ll 1$, where $Kn = h/L$ is the Knudsen number, both effective thermal conductivities, Eq.(48), tend to the bulk value $\lambda$. As the film thickness $L$ decreases or $Kn$ increases, the effective thermal conductivity $\lambda_+^{eff}$ decreases, whereas the "internal" effective thermal conductivity $\lambda_-^{eff}$ increases and tends to infinity in the ballistic regime. The physical interpretation of this fact is that in this regime the temperature gradient tends to zero, while the heat flux through the film has a finite value. To fulfil the Fourier law with a finite value of the heat flux and vanishing temperature gradient, the "internal" effective thermal conductivity $\lambda_-^{eff}$ tends to infinity. Thus, the deviation of $\lambda_-^{eff}$ and $\lambda_+^{eff}$ from their bulk value $\lambda$ with increasing $Kn$ implies that the steady-state heat transport across the thin film occurs under local nonequilibrium conditions. It implies that when $Kn \geq 1$ the classical (local equilibrium) definition of temperature is not valid even for the steady-state regimes. So the concept of the effective temperature should be used.

One might argue, however, that the mean free path $h$ in the DVM can at the most be equal to $L$, that is, in the boundary scattering regime, and therefore the smallest value of $L/h$ is unity [24,26]. It is important to note that the mean free path $h$ is a statistical quantity and can be physically interpreted by the relation $p = \exp(-x/h)$. Here $p$ is the probability that a particle would travel a distance $x$ without undergoing a collision. Therefore it is possible to have $h \gg L$, which means that the probability of a phonon, emerging from one boundary and not being scattering until it reaches the other boundary is $\exp(-L/h)$ [24].

## 2. Effective temperature

The DVM predicts the following expression for the heat flux $q$ across a thing film in steady state regime [26]

$$q = \frac{Cv\Delta T}{2(1+L/h)} \qquad (49)$$

$\Delta T = T_1^R - T_2^R$ is the temperature difference between the thermal reservoirs, $T_i^R$ is the temperature of the thermal reservoir $i$ ($i$=1,2). Using Eqs.(32) and (49), we obtain the effective temperature in a thin film as follows

$$\theta = T\left(1 - \left(\frac{1}{\beta(1+L/h)}\right)^2\right)^{1/2} \qquad (50)$$

where $\beta = 2T/\Delta T$.

Fig.7 shows the nondimensional effective temperature $\theta/T$, Eq.(), versus the Knudsen number $Kn = h/L$. In the Fourier limit $Kn \to 0$ (or $L \gg h$), heat conduction occurs under local equilibrium conditions with $\theta \to T$ (see Fig.7). As $Kn$ increases, there is a substantial deviation from local equilibrium, which is manifested by decreasing effective temperature $\theta/T$ (see Fig.7). This implies that under local nonequilibrium conditions ($Kn \sim 1$) the steady state heat transport through a thin film includes both diffusive (disordered) and ballistic (ordered) modes. The diffusive mode is characterized by the effective temperature $\theta$. The behavior of the effective temperature $\theta$, Eq.(50), corresponds to the behavior of the EIT nonequilibrium temperature in an ideal gas under Couette flow [6,21].

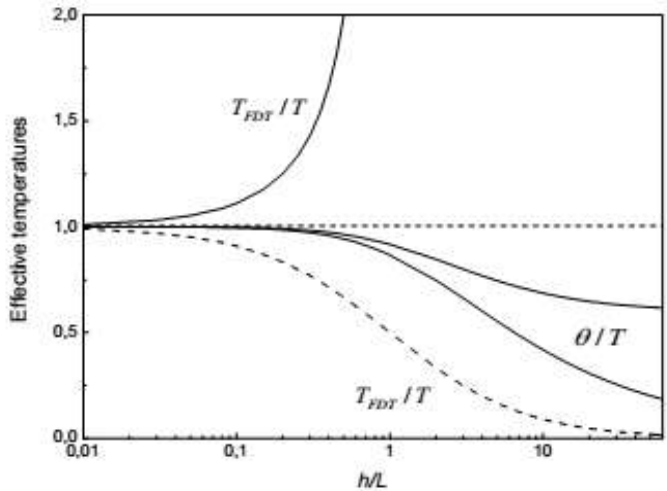

**Fig.7.** Nondimensional temperatures in a thin film as functions of the Knudsen number $Kn = h/L$. The temperatures shown are: the non-equilibrium temperature $\theta/T < 1$ for two different values of $\beta$ - solid lines; the FDT temperature $T_{FDT}/T > 0$, defined through the effective thermal conductivity $\lambda_-^{eff}$ - solid line; the FDT temperature $T_{FDT}/T$, defined through the effective thermal conductivity $\lambda_+^{eff}$ - dashed line.

In the ballistic limit $Kn \to \infty$, $\theta = (T_1^R T_2^R)^{1/2}$, while $T = (T_1^R + T_2^R)/2$. If, for example, $T_2^R \to 0$, then $\theta \to 0$, whereas $T = T_1^R/2 > 0$. This case corresponds to the totally ordered situation when all the heat carriers move in the same direction. Thus, this example demonstrates that the effective temperature $\theta$ can be significantly different from the kinetic temperature $T$ (see Eq.(50) and Fig.7) in nonequilibrium steady-states. At first sight this result seems to be surprising because in steady-state the MFL, Eq.(1), and HHCE, Eq.(2), reduces to the classical local equilibrium FL and PHCE, respectively. However, as demonstrates Eq.(32), there is no difference between $T$ and $\theta$ only in global equilibrium with $q=0$, whereas the difference always exists even in steady state and even for small value of the heat flux $q$ when an assumption of the local equilibrium is valid. Moreover, it should be kept in mind that although the MFL and the HHCE reduce to the FL and the PHCE in steady state, the local nonequilibrium boundary conditions differ from that in local equilibrium even at the steady state [26].

## 3. FDT temperature

Another effective non-equilibrium temperature may be defined from the FDT [3,5,6,17,31,40]. The well-known Einstein relation $D = k_B T \mu$, where $k_B$ is the Boltzmann constant, $\mu$ is the mobility, $D$ is the diffusion coefficient, expresses the relation between fluctuation ($D$) and response ($\mu$). When manifested in a more general manner, this relation is called fluctuation-dissipation theorem. The FDT states a general relationship between the response of a given system to an external disturbance and the internal fluctuations of the system in equilibrium. This relationship contains the temperature and is central in thermodynamics. However, when a system is out of equilibrium, the theorem breaks down and an extension of the theorem must be made. There is growing evidence that a modified form of the FDT with corresponding effective temperature holds out of equilibrium in a wide , of conditions for example, in glassy systems in the ageing regime, jammed granular media, and non-equilibrium steady states in models of driven and active matter [5,6,17,31,40]. Following Palacci et al. [31], we introduce the FDT temperature $T_{FDT}$ using the effective transport coefficient $\lambda_{\pm}^{eff}$, Eq.(48), which results in

$$T_{FDT\pm}/T = \lambda_{\pm}^{eff}/\lambda \qquad (51)$$

The ratios $T_{FDT\pm}/T$ are also shown in Fig. as functions of the Knudsen number $Kn = h/L$. In contrast to $\theta$, the FDT temperature $T_{FDT-}$, which is based on the effective thermal conductivity $\lambda_{-}$, increases with increasing deviation from equilibrium (increasing $Kn$). The increase of $\lambda_{-}$ compensates the decrease of the temperature gradient inside the film when $Kn \to \infty$. The behavior of $T_{FDT-}$, Eq.(51), and corresponds to the behavior of the FDT temperature in an ideal gas under Couette flow [5,6,21], in an active colloidal suspension under gravity field [31], and in an ensemble of interacting self-propelled semi-flexible polymers [40].

C. Effective temperature in shock wave – comparison with MD simulation

The shock-wave propagation occurs under strong nonequilibrium conditions because the shock fronts is highly localized in both distance (a few interatomic spacings) and time (a few mean collision times) [30]. Due to the far-from-equilibrium nature of the shock wave the average kinetic temperature $T_k$ is defined it in terms of the local peculiar kinetic energy; hence $T$ is one-third the trace of the kinetic temperature tensor [30]. In the shock front, the kinetic temperature component in the direction of shock propagation, $T_{xx}$, is higher than the transverse components, $T_{yy}$ and $T_{zz}$, which are equal to each other by symmetry. Therefore $T_k$ is also always lower than $T_{xx}$, except at equilibrium, which occurs long before the shock has arrived and long afterwards, when equipartition holds. Moreover, $T_{xx}$ shows a distinct peak near the center of the shock front, and this disequilibrium is due to collisions in the shock compression process [30]. The temperatures $T_{xx}$ and $T_k$ in the work of Holian et al. [30] correspond to $T$ and $\theta$ in the present model, respectively. To compare the MD results with the present model, we take the data for $q(x)$ and $\theta(x)$ from Fig.3 in Ref.[30] and then calculate $T$ (analog to $T_{xx}$) from Eq.(32). All the functions were normalized to the corresponding equilibrium values at $x \to \infty$ taken from [30], so we do not need to know the heat capacity and phonon speed to calculate $T$ from Eq.(32). Fig.8 shows the effective (average) temperature $\theta$ (dashed curve), the longitudinal component of temperature in the shock-wave direction $T$ form Eq.(32) (solid curve), the MD data for $T_{xx}$ from Ref.[30] (solid circles), and the heat-flux $q$ (dash-dotted curve) as functions of coordinate $x$ for a strong shockwave in the Lennard-Jones dense fluid ($x=0$ is the wave front).

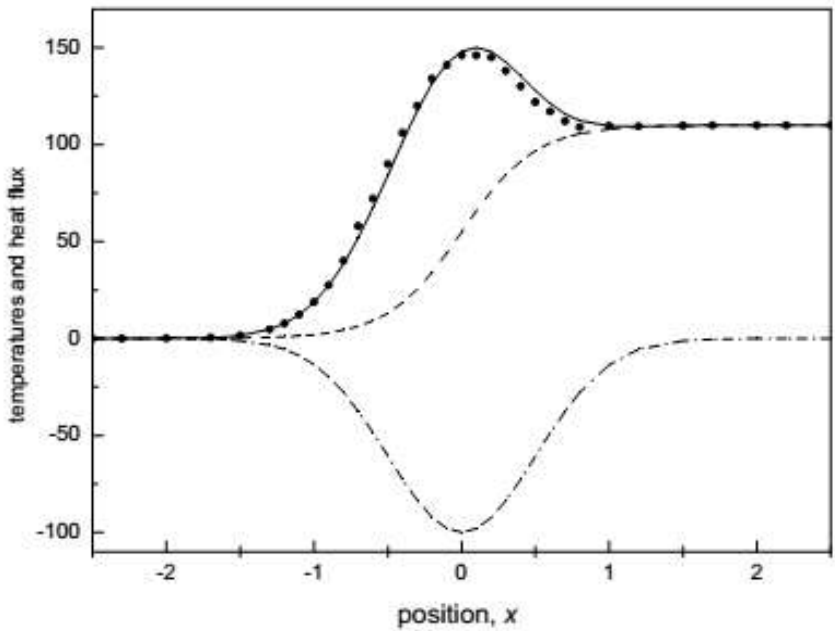

**Fig.8.** Nondimensional temperatures and heat flux distributions as functions of coordinate $x$ for a strong shockwave ($x=0$ is the wave front). Solid line is the longitudinal component of the temperature in the direction of the shockwave $T$ (or $T_{xx}$ in terms of Ref.[30]) calculated from Eq.(32); solid circles is the nonequilibrium molecular dynamics simulation data for $T_{xx}$ [30]; dashed line and dash-dotted lines are the average (or effective) temperature $\theta$ and the heat flux $q$, respectively, taken from Ref.[30].

Comparison of the behavior of $T$ calculated from the present model (solid curve in Fig.8) and the nonequilibrium MD data for $T_{xx}$ (solid circles) taken from Ref.[30] demonstrates good agreement. Thus, the present model, Eq.(32), correctly describes the relationship between the temperatures, $T_{xx}$, $T_k$, and the heat flux $q$ in the front of the strong shock waves. Note that a distinct peak of the longitudinal temperature near the wave front due to nonequilibrium effects has been predicted earlier around a fast-moving heat source [12].

## V. CONCLUSION

Random walk approach with an assumption of a finite value of heat (mass) curriers velocity leads to the HHCE, Eq.(2), and the MFL, Eq.(1) [1,2]. The result also corresponds to the DVM with the wave law of continuum limit [12,13,26,34-36] and to the BTE with the single relaxation time approximation [6,10,18,24,37,38,42]. The HHCE and the MFL describe the space-time evolution of the local nonequilibrium system with the kinetic temperature $T$, which characterizes the local energy density, i.e. $T$ corresponds to the equilibrium temperature of the same system with the same local energy. In other words, if a local volume element of the nonequilibrium system is suddenly isolated, i.e. bounded by adiabatic and rigid walls, and allowed to relax to equilibrium, after equilibration the system temperature will be $T$. In 1D the local kinetic temperature $T$ is an average of the kinetic temperatures of the heat carries, $T_1$ and $T_2$, moving in opposite directions. If the same local volume equilibrates reversibly, i.e. while producing work, after equilibration its temperature will be $\theta = (T_1 T_2)^{1/2}$. The effective temperature $\theta$ characterizes the thermal (equilibrated) fraction of the energy density under nonequilibrium conditions and can serve as a criterion for thermalization. The effective temperature $\theta$ depends on the heat flux $q$ and is governed by the non-linear relation (Eq.(32)). When the heat flux $q$ tends to its upper limit $|q| \to q^{\max}$, the nonequilibrium approach predicts a third-law-like behavior in terms of the corresponding nonequilibrium quantities, namely, $\theta \to 0$, $S \to 0$, $C_{neq} \to 0$, even at a non-zero value of $T$.

The approach provides a further basis for the understanding of the effective temperature and entropy in a wide range of nonequilibrium systems, from graphene-like materials to active matter in biology. However, a comprehensive formulation of the concepts of temperature and entropy out of equilibrium for more complex systems, particularly in the quantum limit, is still an open problem and requires additional research.

## ACKNOWLEDGMENTS

The reported study was partially supported by the Russian Foundation for Basic Research, Research Projects No. 16-03-00011.

**References.**
[1] V.A. Fock, Transactions of the Optical Institute in Leningrad **4**, 1 (1926).
[2] B.I. Davydov, Dokl. Akad. Nauk SSSR **2**, 474 (1935) [C. R. Acad. Sci. URSS **2**, 476 (1935)]
[3] L.D. Landau and E.M. Lifshitz, *Statistical Physics* ( Pergamon Press, Oxford, 1970).
[4] L.D. Landau and E.M. Lifshitz, *Fluid Mechanics* ( Pergamon Press, Oxford, 1966).
[5] J. Casas-Vázquez and D. Jou, Rep. Prog. Phys. **66**, 1937 (2003).
[6] D. Jou, J. Casas-Vázquez, and G. Lebon, *Extended Irreversible Thermodynamics* (Springer, Berlin, 2010).
[7] P. Ván and T. Fülöp, Ann. Phys. (Berlin) **524**, 470 (2012).
[8] R.E. Nettleton and S.L. Sobolev, J. Non-Equilibr. Thermodyn. **20**, 205 (1995); **20**, 297 (1995); **21**, 1 (1996).

[9] D. Jou and L. Restuccia, Physica A **460**, 246 (2016); L. Restuccia, Commun. Appl. Ind. Math. **7**, 81 (2016).

[10] Y. Guo and M. Wang, Phys. Rep. **595**, 1 (2015).

[11] V.A. Cimmelli, J. Non-Equilib. Thermodyn. **34**, 299 (2009).

[12] S.L. Sobolev, Usp. Fiz. Nauk. **161**, 5 (1991) [Sov. Phys. Usp. **34**, 217 (1991)].

[13] S.L. Sobolev, Usp. Fiz. Nauk. **167**, 1095 (1997) [Phys. – Usp. **40**, 1043 (1997)].

[14] D.D. Joseph and L. Preziosi, Rev. Mod. Phys. **61**, 41 (1989); **62**, 375 (1990).

[15] H.G. Weiss, Physica A **311**, 381 (2002).

[16] D.G. Cahill, W.K. Ford, K.E. Goodson, G.D. Mahan, A. Majumdar, H.J. Maris, R. Merlin, and S.R. Phillpot, J. Appl. Phys. **93**, 793 (2003).

[17] J. Kurchan, Nature, **433**, 222 (2005).

[18] S. Sinha and K. E. Goodson, Int. J. Multiscale Comput. Eng. **3**, 107 (2005).

[19] S.L. Sobolev, Mater. Sci. Techn. **31**, 1607 (2015).

[20] M.E. Siemens, Q. Li, R. Yang, K.A. Nelson, E.H. Anderson, M.M. Murnane, and H.C. Kapteyn, Nat. Mater. 9, **26** (2010).

[21] M. Criado-Sanchoa, D. Jou, and J. Casas-Vázquez, Phys. Lett. A **350**, 339 (2006).

[22] G. Chen, J. Heat Transf. **118**, 539 (1996).

[23] G. Chen, J. Heat Trans. **124**, 328 (2002).

[24] A. Majumdar, J. Heat Transf. **115**, 7 (1993).

[25] S.L. Sobolev, Phys. Rev. E **55**, 6845 (1997).

[26] S.L. Sobolev, Int. J. Heat Mass Trans. **108**, 933 (2017).

[27] J. Camacho, Phys. Rev. E **51,** 220 (1995).

[28] E. Kronberg, A.H. Benneker, and K.R. Westerterp, Int. J. Heat Mass Transf. **41**, 127 (1998).

[29] P.K. Patra and R.C. Batra, Phys. Rev. E **95**, 013302 (2017).

[30] B.L. Holian, M. Mareschal, and R. Ravelo, Phys. Rev. E **83**, 026703 (2011).

[31] J. Palacci, C. Cottin-Bizonne, C. Ybert, and L. Bocquet, Phys. Rev. Lett. **105**, 088304 (2010).

[32] T. Lu, J. Hasty, and P.G. Wolynes, Biophysical J. **91**, 84 (2006).

[33] C. Essex, R. McKitrick, and B. Andresen, J. Non-Equilibr. Thermodyn. **32**, 1 (2007).

[34] S.L. Sobolev, J. Phys. III France **3**, 2261 (1993).

[35] S.L. Sobolev, Int. J. Heat Mass Transf. **71**, 295 (2014).

[36] S.L. Sobolev, Phys. Rev. E **50**, 3255 (1994).

[37] R.A. Escobar, S.S. Ghai, M.S. Jhon, and C.H. Amon, Int. J. Heat Mass Trans. **49**, 97 (2006).

[38] S. Pisipati, J. Geer, B. Sammakia, and B.T. Murray, Int. J. Heat Mass Trans. **54**, 3406 (2011).

[39] S. Voltz, J.-B. Saulnier, M. Lallemand, B. Perrin, P. Depondt, and M. Marescha, Phys. Rev. B **54,** 340 (1996).

[40] C. Bechinger, R. Di Leonardo, H. Löwen, C. Reichhardt, G. Volpe, and G. Volpe, Rev. Mod. Phys. **88**, 045006 (2016); L. F. Cugliandolo, J. Phys. A: Math. Theor. **44**, 483001 (2011).

[41] Y. Dong, B.Y. Cao, and Z.Y. Guo, Phys. Rev. E **87**, 032150 (2013).

[42] J. Xu and X. Wang, Physica B **351**, 213 (2004).

[43] S. Marianer and B.I. Shklovskii, Phys. Rev. B **46**, 13100 (1992).

[44] S.D. Baranovskii, B. Cleve, R. Hess, and P. Thomas, J. Non-Cryst. Solids, **164–166**, 437 (1993).

[45] C.E. Nebel, R.A. Street, N.M. Johnson, and C.C. Tsai, Phys. Rev. B **46**, 6803 (1992).

[46] G. Liu and H.H. Soonpaa, Phys. Rev. B **48**, 5682 (1993).

[47] D. Loi, S. Mossa, and L. F. Cugliandolo, Soft Matter, **7**, 3726 (2011); **7**, 10193 (2011).

[48] A. Pachoud, M. Jaiswal, Yu. Wang, B.-H. Hong, J.-H. Ahn, K. P. Loh, and B. Ozyilmaz, Sci. Rep. **3**, 3404 (2013); DOI:10.1038/srep03404 (2013).

[49] S.L. Sobolev, Phys. Lett. A (accepted).